\documentclass[%
 reprint,
 amsmath,amssymb,
 aps,prd
]{revtex4-2}

\usepackage{graphicx}
\usepackage{dcolumn}
\usepackage{bm}
\usepackage{makecell}
\usepackage{nicematrix}
\usepackage[colorlinks=true,linkcolor=blue,citecolor=blue,urlcolor=blue]{hyperref}

\newcommand{\finesse}{\textsc{Finesse}}

\begin{document}


\preprint{APS/123-QED}

\title{Angular control noise in Advanced Virgo and implications for the Einstein Telescope}

\author{Riccardo Maggiore}
\affiliation{Department of Physics and Astronomy, VU Amsterdam; De Boelelaan 1081, 1081, HV, Amsterdam, The Netherlands}
\affiliation{Nikhef; Science Park 105, 1098, XG Amsterdam, The Netherlands}
\email{r.maggiore@nikhef.nl}

\author{Paolo Ruggi}
\affiliation{European Gravitational Observatory; Via E. Amaldi, 5, 56021 Cascina (PI), Italy}
\email{ruggi@ego-gw.it}

\author{Andreas Freise}
\affiliation{Department of Physics and Astronomy, VU Amsterdam; De Boelelaan 1081, 1081, HV, Amsterdam, The Netherlands}
\affiliation{Nikhef; Science Park 105, 1098, XG Amsterdam, The Netherlands}
\email{a.freise@nikhef.nl}

\author{Daniel Brown}
\affiliation{Department of Physics, The University of Adelaide, South Australia, 5005, Australia}
\affiliation{The Institute of Photonics and Advanced Sensing (IPAS), The University of Adelaide, South Australia, 5005, Australia}

\author{Jonathan W. Perry}
\affiliation{Department of Physics and Astronomy, VU Amsterdam; De Boelelaan 1081, 1081, HV, Amsterdam, The Netherlands}
\affiliation{Nikhef; Science Park 105, 1098, XG Amsterdam, The Netherlands}
\email{jperry@nikhef.nl}

\author{Enzo N. Tapia San Martín}
\affiliation{Nikhef; Science Park 105, 1098, XG Amsterdam, The Netherlands}
\email{E.Tapia@nikhef.nl}

\author{Conor M. Mow-Lowry}
\affiliation{Department of Physics and Astronomy, VU Amsterdam; De Boelelaan 1081, 1081, HV, Amsterdam, The Netherlands}
\affiliation{Nikhef; Science Park 105, 1098, XG Amsterdam, The Netherlands}
\email{c.m.mow-lowry@vu.nl}

\author{Maddalena Mantovani}
\affiliation{European Gravitational Observatory; Via E. Amaldi, 5, 56021 Cascina (PI), Italy}
\email{maddalena.mantovani@ego-gw.it}

\author{Julia Casanueva Diaz}
\affiliation{European Gravitational Observatory; Via E. Amaldi, 5, 56021 Cascina (PI), Italy}
\email{julia.casanueva@ego-gw.it}

\author{Diego Bersanetti}
\affiliation{INFN, Sezione di Genova, I-16146 Genova, Italy}

\author{Matteo Tacca}
\affiliation{Nikhef; Science Park 105, 1098, XG Amsterdam, The Netherlands}
\email{m.tacca@nikhef.nl}

\date{\today}


\begin{abstract}
With significantly improved sensitivity, the Einstein Telescope (ET), along with other upcoming gravitational wave detectors, will mark the beginning of precision gravitational wave astronomy. However, the pursuit of surpassing current detector capabilities requires careful consideration of technical constraints inherent in existing designs. The significant improvement of ET lies in the low-frequency range, where it anticipates a one-million-fold increase in sensitivity compared to current detectors. Angular control noise is a primary limitation for LIGO detectors in this frequency range, originating from the need to maintain optical alignment. Given the expected improvements in ET's low-frequency range, precise assessment of angular control noise becomes crucial for achieving target sensitivity. To address this, we developed a model of the angular control system of Advanced Virgo, closely matching experimental data and providing a robust foundation for modeling future-generation detectors. Our model, for the first time, enables replication of the measured coupling level between angle and length. Additionally, our findings confirm that Virgo, unlike LIGO, is not constrained by alignment control noise, even if the detector were operating at full power.
\end{abstract} 

\maketitle

\section{Introduction}
\begin{figure*}[!t]
  \begin{minipage}{0.475\textwidth}
    \centering
    \includegraphics[width=\linewidth]{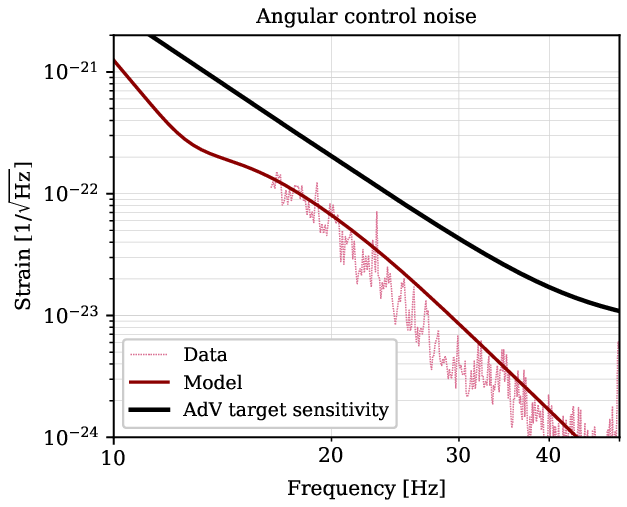}
  \end{minipage}
  \hfill
  \begin{minipage}{0.475\textwidth}
    \centering
    \includegraphics[width=\linewidth]{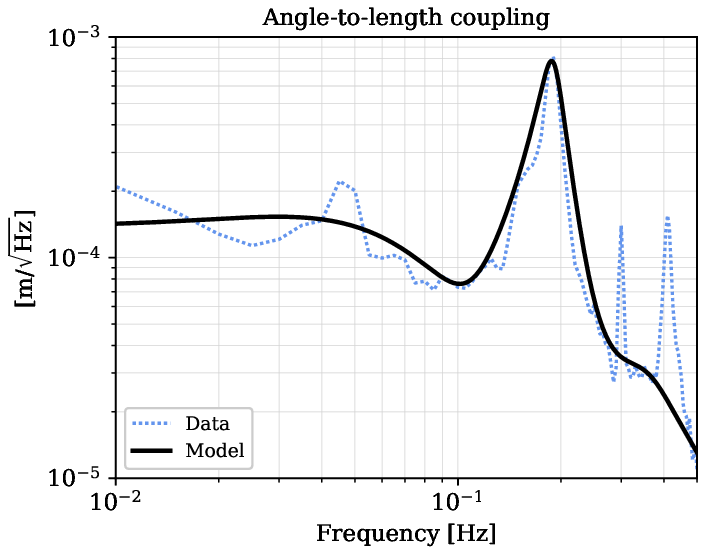}
  \end{minipage}
  \caption{Two examples of the excellent agreement between the angular control noise model developed for Advanced Virgo and experimental data. On the left, the level of noise calculated using the model and that measured experimentally are shown. The experimental data are those presented in~\cite{Bader2021}. The target sensitivity for Advanced Virgo for O3 is shown in black. On the right, the Beam-Spot Motion (BSM) on one of the mirrors is shown. Further clarification is provided later, emphasizing that the BSM is responsible for the angle-to-length coupling.}
  \label{fig:projections_tot}
\end{figure*}

The Einstein Telescope~\cite{Et2020}, along with other forthcoming detectors, will advance gravitational wave detection capabilities to an unprecedented level, enabling the observation of previously undetectable events, such as the mergers of intermediate-mass black holes, and conducting precise tests of Einstein's theory of general relativity under extreme gravitational conditions. Nevertheless, as the objective is to surpass the capabilities of current detectors, it is logical to assume that the technical constraints inherent in existing detectors might hinder the achievement of ET's goals unless the design is carefully crafted to avoid such limitations. The significant improvement that will be brought about by ET lies primarily in the low-frequency range, where an approximately one million-fold increase in sensitivity is expected compared to current detectors. In this frequency range, the primary limitation of LIGO detectors~\cite{Aasi_2015} is due to angular control noise that arises from the need to maintain optical alignment ~\cite{Barsotti_2010, PhysRevD.102.062003, PhysRevD.93.112004}. 
Maintaining the operational integrity of the detector requires a high level of alignment stability between the mirrors. Using feedback control ensures that the remaining rotation of the mirrors remains sufficiently low to meet these requirements. However, this comes at the cost of generating control noise that can couple with the detector's output channel, potentially limiting its performance. Considering the significant expected improvements in this frequency range for ET, it is reasonable to assume that angular control noise may pose a limitation in achieving its target sensitivity. Therefore, it is crucial to be able to evaluate the level of this control noise for ET, and to do this, we needed a precise model. This had not been achieved previously, mainly because there was hope that simplified models would suffice. We took a different approach and developed a more detailed model of the angular control system for the Advanced Virgo detector (AdV)~\cite{Acernese_2015}. Our model is closely aligned with experimental data, providing a robust foundation for simulating future-generation detectors. Remarkably, we have successfully replicated, for the first time, the measured coupling level between angle and length; see Fig.~\ref{fig:projections_tot}.\\

Incidentally, our model confirmed that Virgo, unlike LIGO, is not constrained by alignment control noise, as previously shown in~\cite{Bader2021}. Through the validated model, we show that this limitation would not apply even if the detector was operating at full power.\\

In Section~\ref{sec:asc_1}, we begin by explaining the fundamentals of angular control noise and how mirror rotations can lead to a length signal. We will clarify that assessing the amount of angular noise requires understanding the dynamics of the entire control system. Therefore, we begin the description of angular control loops in Section~\ref{sec:asc_2}, starting with their fundamental component, the plant. Specifically, we present an analytical model to describe the influence of radiation pressure on the suspension system of optics. In the same section, we provide a concise overview of how the entire control system model was assembled. More details about the plant can be found in Appendix~\ref{appendix:modes}, while the architecture of the control system can be found in Appendix~\ref{appendix:control}. In Section~\ref{sec:asc_3}, we introduce the external disturbances that enter the angular control loops. These disturbances act as a source of angular control noise. For a more detailed description of all external disturbances, see Appendix~\ref{appendix:noise}. In Section~\ref{sec:asc_4},  we briefly explain how we calculated the level of angular control noise coupled to the detector output and show the noise projection in the scenario where the detector was operated at full power. A more detailed explanation of the methodology behind generating noise projections can be found in Appendix~\ref{appendix:proj}. Lastly, we conclude by presenting the same noise projection in the case of the Einstein Telescope in Section~\ref{sec:asc_5}. In Appendix~\ref{appendix:ETmech}, we provide details of the suspension mechanics that were taken into account to produce the estimate of the noise level for ET.

\section{\label{sec:asc_1}Angular control noise: the coupling chain}
\begin{figure}[!b]
	\centering
	\includegraphics[width=\linewidth]{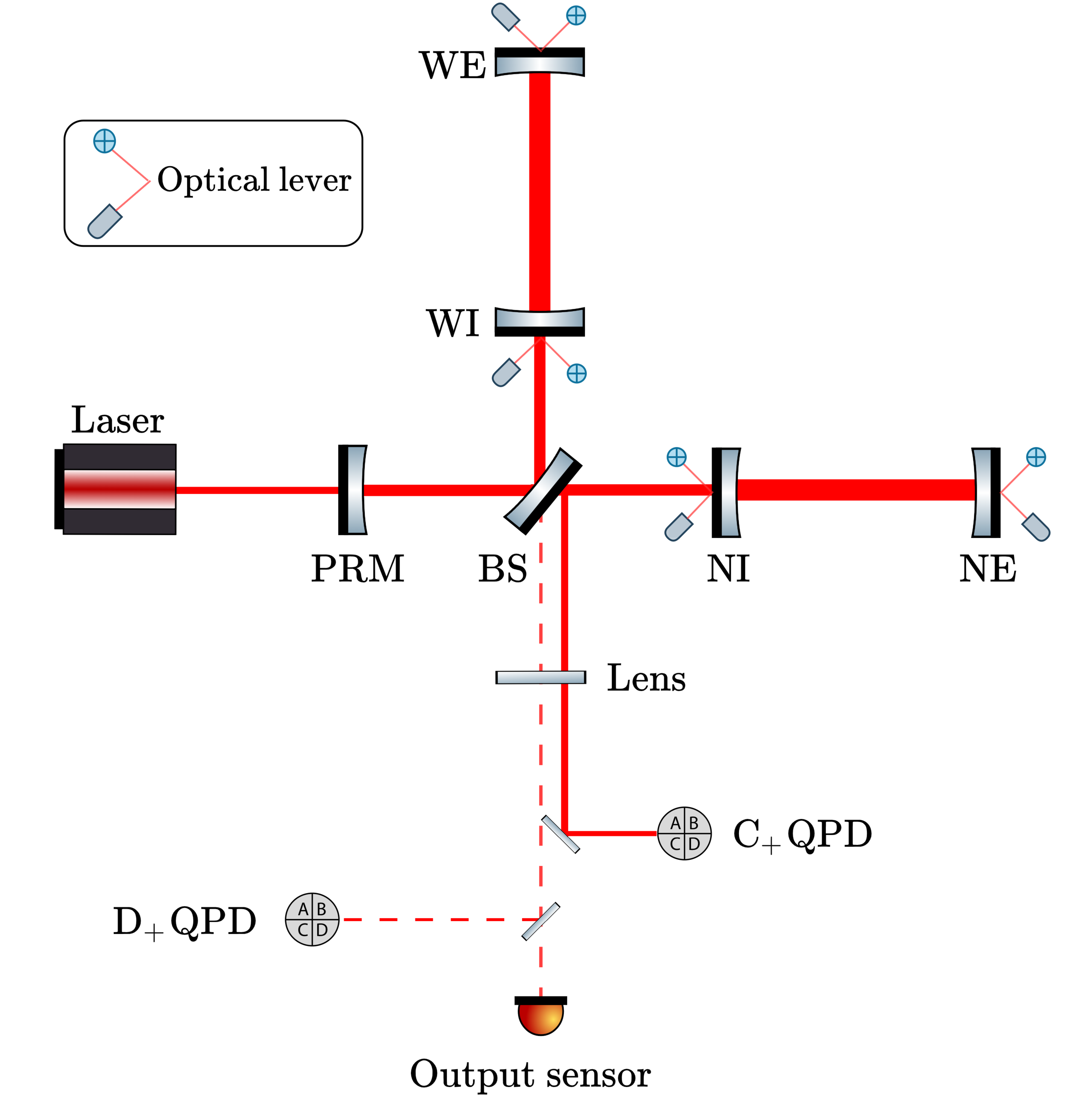}
	\caption{\label{fig:ifo_scheme}The diagram shows Advanced Virgo's optical configuration during the O3 run. The main optics consist of input test masses (NI, WI), end test masses (NE, WE), the Power Recycling Mirror (PRM), the beam splitter (BS), and a lens that was used as a placeholder for the signal recycling mirror. The diagram also indicates the positions of the alignment sensors, namely quadrant photodetectors (QPD) and optical levers (OpL).}
\end{figure}

A diagram of the interferometric detector is shown in Fig.~\ref{fig:ifo_scheme}. The detector configuration that we modeled is known as Advanced Virgo and was utilized during the third observational run O3. As of now, the current commissioning phase for Virgo in preparation for the fourth observational run is still ongoing. We believed that modeling a stable system that is not subject to changes was the right environment to develop our model.\\

\begin{figure*}[!t]
	\centering
	\includegraphics[width=0.72\linewidth]{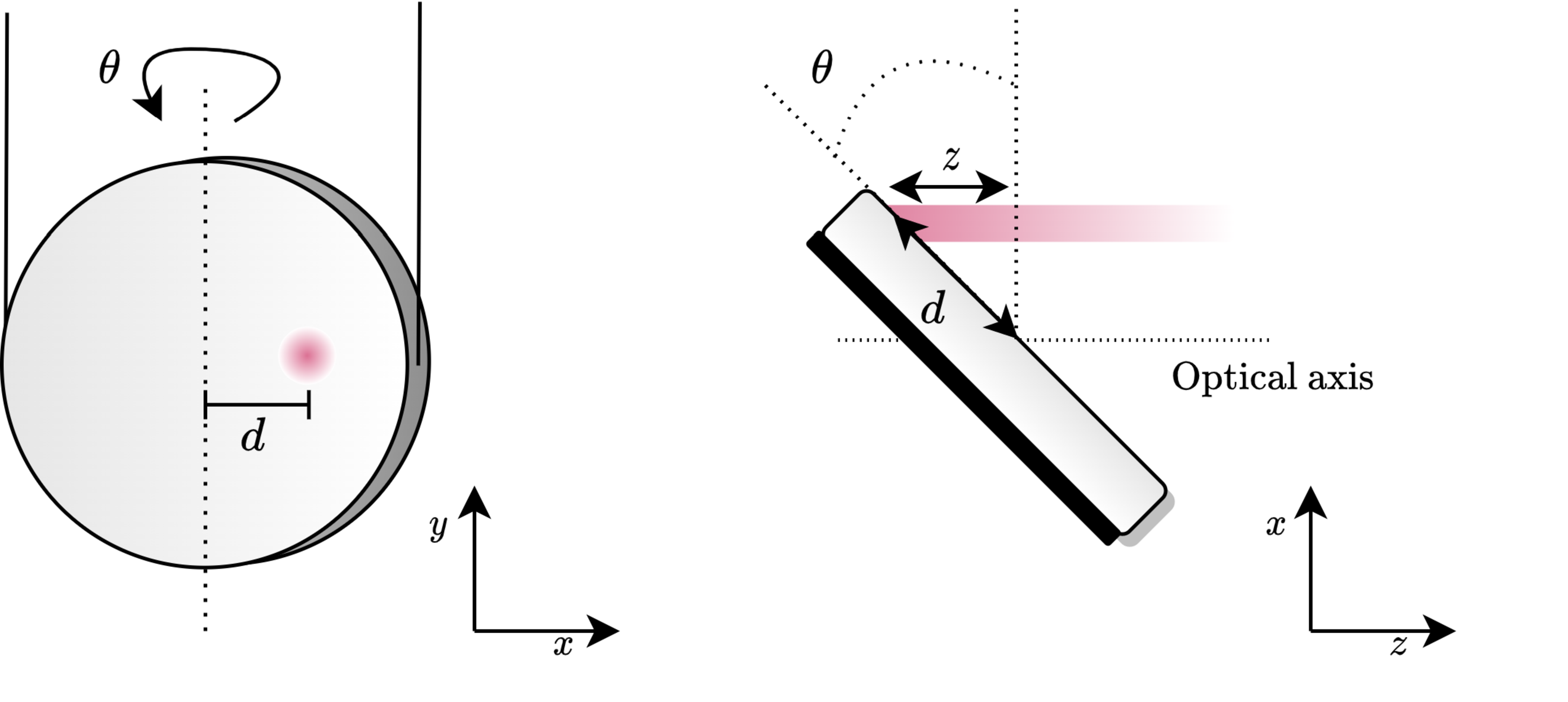}
	\caption{\label{fig:asc_mirror}Illustration in the x-y plane (on the left) and in the z-x plane (on the right) of a beam incident on a mirror, with an offset $d$ with respect to the rotation axis of the optics. The figure aims to show how a mirror rotation can alter the optical path of a beam due to the offset of the beam.}
\end{figure*}

When a gravitational wave passes through the detector, it induces a differential change in the length of the detector arm cavities. Similarly, any rotation of the main optics can also lead to a change in length, which can create noise that is indistinguishable from a gravitational wave. As illustrated in Fig.~\ref{fig:asc_mirror}, this angle-length coupling can be easily understood from a geometric point of view. If the incident beam in one of the mirrors is offset by a value $d$, the rotation of the mirror by a small angle $\theta$ results in a corresponding change in length given by:
\begin{equation}\label{eq:angle2length0}
	z = d \, \theta
\end{equation}
Eq.~\ref{eq:angle2length0} is simply an angle times a distance, which gives the change in the optical path experienced by the beam. The beam on the mirrors is off-centered due to the rotation of the optics. To illustrate the concept, let us consider the case of a single arm cavity with an Input Test Mass (ITM) and an End Test Mass (ETM). As explained in~\cite{siegman1986lasers}, the beam spot motion on the Test Masses (TMs) is related to their rotation by the following relationship.
\begin{equation}\label{eq:bsm}
    [d_{\textnormal{ITM}} \; d_{\textnormal{ETM}}]^{T} = D \cdot [\theta_{\textnormal{ITM}} \; \theta_{\textnormal{ETM}}]^{T}
\end{equation}
with
\begin{equation}\label{eq:beam_offset}
    D =\frac{L}{1-g_{\textnormal{ITM}}\;g_{\textnormal{ETM}}}\;\begin{bNiceMatrix} 
        g_{\textnormal{ITM}} & -1 \\
        -1 & g_{\textnormal{ETM}}\\
    \end{bNiceMatrix}
\end{equation}
The matrix $D$ connects the rotation of the TMs, $\theta_{\textnormal{ITM}}$ and $\theta_{\textnormal{ETM}}$, to the displacement of the beam spot on each respective TM, $d_{\textnormal{ITM}}$ and $d_{\textnormal{ETM}}$. For each TM, the parameter $g_{\textnormal{TM}}$ depends on the geometry of the cavity as:
\begin{equation}
    g_{\textnormal{TM}} = 1 - L/R_{\textnormal{TM}}
\end{equation}

Eq.~\ref{eq:bsm} shows that the cause of angular noise is the rotation of the optics, which causes the beam to become off center. Now, the last thing left to understand is why the optics undergo these movements.\\
External disturbances, denoted by $\eta$, enter the control loops, causing the movement of the mirrors. Therefore, Eq.~\ref{eq:angle2length0} can be refined to include more specific dependencies as follows:
\begin{equation}\label{eq:angle2length}
z(\theta(\eta)) = d(\theta(\eta)) \ast \theta(\eta)
\end{equation}

The entire process unfolds as follows: external noise induces mirror movement $\theta(\eta)$, which, in turn, triggers the BSM $d(\theta(\eta))$; within a bilinear process, the BSM modulates the angular motion, ultimately producing a length signal. Eq.~\ref{eq:angle2length} shows how an angle can generate a length signal when considering a single mirror. However, when dealing with the entire detector, which comprises multiple mirrors, assessing the amount of angular control noise coupled to the output channel becomes more complex than the expression provided in Eq.~\ref{eq:angle2length}. Nevertheless, that equation is adequate to grasp that two essential components are required to evaluate the amount of angular noise: identifying and characterizing the input noises that cause the motion of the optics, and a control loop model that relates mirror motion to the set of input noises. This is because the dynamics of the optics is shaped by the entire control system. In this framework, we will begin by describing the core component of the Alignment Sensing and Control (ASC) system in the next section, which is its plant.

\section{\label{sec:asc_2}The opto-mechanical plant and degrees of freedom}
The detector requires precise alignment of the optics in both the yaw (horizontal) and pitch (vertical) axes. Our analysis specifically addresses yaw, which is sufficient to establish a robust methodology for future detector modeling. Henceforth, we will imply that all Degrees of Freedom (DoFs) we define are referenced to the horizontal rotation axis. Expanding this analysis to include pitch requires a slightly more intricate mechanical description. However, extending this modeling to pitch remains relatively straightforward.\\
Furthermore, we will only consider the DoFs of the arm cavities because their rotation directly influences the cavity length, thus directly coupling with the detector output. This is not the case for other optics, whose coupling with the output is indirect. For example, the rotation of the PRM, when there is a misalignment of the beam, affects the length of the power recycling cavity, not that of the arms.\\ 

In this section, we illustrate the effect of radiation pressure on the mechanics that suspend the optics. The description of the mechanical system that follows aims to be a more concise version of the one in~\cite{maggiore2024}.\\
As elaborated later in this section, the two arm cavities operate as mechanically independent systems. Consequently, our discussion of the effects of radiation pressure begins by focusing on a single arm cavity before extending our analysis to both cavities.\\

To provide isolation from ground vibrations, the TMs in each arm cavities are suspended using a multistage seismic isolator called the superattenuator, as detailed in~\cite{BCaron_1997}. The lower part of the suspension consists of a steering element, the marionette, from which the suspended optics hang~\cite{doi:10.1063/1.1463717}. The actuators used in the control scheme are attached to the lower stages of the suspension chain. Therefore, considering the mechanical response of the suspension is necessary to describe the optics dynamics. During the O3 run, the control scheme used only the actuators anchored to the marionette. Hence, we simplify our analysis, assuming that each TM is suspended solely by the last two stages of suspension, as the ASC system only involves these.\\

\begin{figure}[!b]
	\centering
	\includegraphics[width=\linewidth]{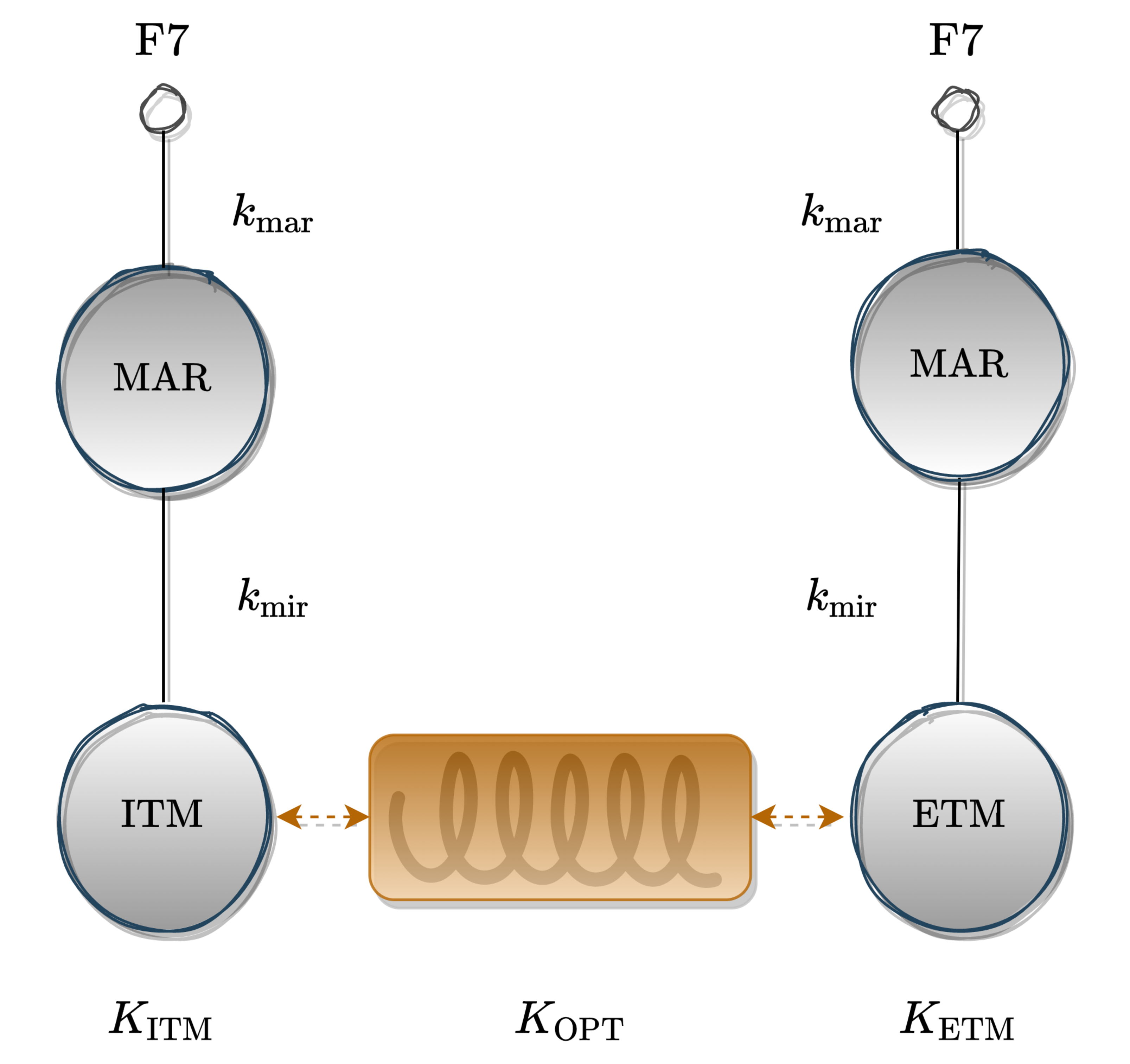}
	\caption{\label{fig:mechanicsRP} An illustration of the optomechanical system for each arm cavity. Both TMs are suspended using their respective two-stage suspension. The second to last stage is called the MARionette (MAR). Due to the impact of radiation pressure, the two TMs behave as if connected by a spring, effectively coupling the two suspension systems. The stiffness term $K_{\textnormal{OPT}}$ accounts for the radiation pressure effect. The diagram shows the suspension point of the marionette, situated in the upper stage of the suspension chain known as Filter 7 (F7).}
\end{figure}

In a gravitational wave detector, a substantial amount of optical power is stored within each arm cavity. This leads to a non-negligible impact of radiation pressure on the mechanics suspending the optics. For each arm cavity, the role of radiation pressure is to couple the motion of the TMs. As explained in Section~\ref{sec:asc_1}, the rotation of one of the TMs offsets the beam on both mirrors. The beam exerts a force on both optics through radiation pressure at a point now offset from the center of rotation, resulting in a torque applied to both TMs. Therefore, the rotation of one TM can induce the motion of the other. On the other hand, in the absence of radiation pressure, the suspensions of each TM operate as mechanically independent systems, with the rotation of one optic not inducing torque on the other optic. When there is radiation pressure, the system cannot be characterized as two independent suspension systems. Instead, it operates as a unified mechanical system, often referred to as an optomechanical system, underscoring the significant role of optics. Fig.~\ref{fig:mechanicsRP} illustrates a schematic of the optomechanical system, and its dynamics can be described starting from the equations of a simple torsional pendulum as
\begin{equation}\label{eq:asc_motion}
I \ddot{\theta}(t) - K \theta(t) = T(t)
\end{equation}
$T$ represents the driving torque. $\theta$ denotes the following vector of DoFs:
\begin{equation*}
    \theta = [\theta_{\text{mar, ITM}}\;\; \theta_{\text{mir, ITM}}\;\;\theta_{\text{mar, ETM}}\;\; \theta_{\text{mir, ETM}}]^{T}
\end{equation*}
From now on, we will use the subscript \textit{mar} to denote the ``marionette'' stage and \textit{mir} for the final stage, which is the one supporting the ``mirror''. The ITM and ETM subscripts serve as reference elements in the equation corresponding to the ITM or ETM suspension. For example, $\theta_{\text{mir, ITM}}$ represents the motion of the mirror stage of the ITM suspension. $I$ corresponds to the inertia matrix, which is as follows:
\begin{equation*}
    I = \textnormal{diag}(I_{\text{mar, ITM}}\;\; I_{\text{mir, ITM}}\;\; I_{\text{mar, ETM}}\;\; I_{\text{mir, ETM}})
\end{equation*}
$K$ represents the stiffness matrix. To isolate the contribution to stiffness from mechanics and from optics, we write $K$ as follows:
\begin{equation}
    K = K_{\textnormal{M}} + K_{\textnormal{OPT}}
\end{equation}
The matrix $K_{\textnormal{M}}$ is a stiffness matrix of purely mechanical nature, which we express as follows:
\begin{equation}\label{eq:asc_mecc_stiff}
    K_{\textnormal{M}} =
    {\small
    \begin{bNiceMatrix} 
        k_{\text{mar}}+k_{\text{mir}} & -k_{\text{mir}} & 0 & 0\\
        -k_{\text{mir}} & \;\;\;k_{\text{mir}} & 0 & 0\\
        0 & 0 & k_{\text{mar}}+k_{\text{mir}} & -k_{\text{mir}}\\
        0 & 0  & -k_{\text{mir}} & \;\;\;k_{\text{mir}}
    \end{bNiceMatrix}
    }
\end{equation}
\begin{figure*}[!t]
	\centering
	\includegraphics[width=0.95\linewidth]{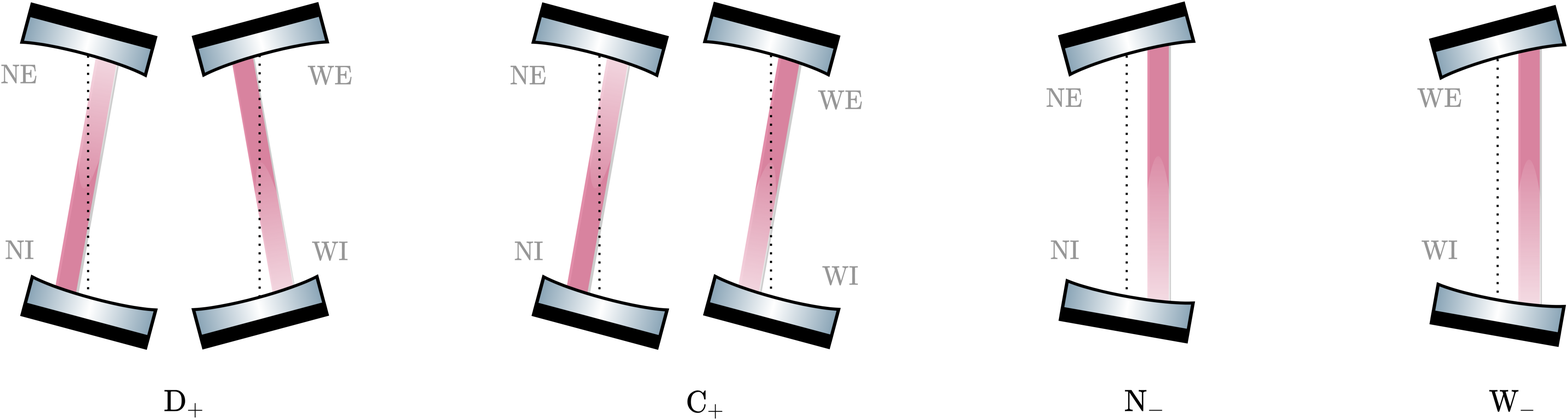}
	\caption{\label{fig:asc_dofs}Illustration of DoFs under control. $\textnormal{D}_+$ stands for Differential Plus, $\textnormal{C}_+$ for Common Plus, $\textnormal{N}_-$ for North Minus, and $\textnormal{W}_-$ for West Minus. Respectively, they include differential tilting of the beam between the arm cavities, common tilting of the beam between the arm cavities, shifting of the beam in the north arm cavity, and shifting of the beam in the west arm cavity. We will refer to the pair of DoFs $\textnormal{D}_+$ and $\textnormal{C}_+$ as ``hard modes'', and the pair  $\textnormal{N}_-$ and  $\textnormal{W}_-$ as ``soft modes''. For more details on the reason for this nomenclature, see Appendix~\ref{appendix:modes}. The detector requires precise alignment of the optics along both the yaw and pitch axes. Our analysis focuses specifically on the yaw axis.}
\end{figure*}
The stiffness of each stage is denoted by $k_{\text{mar}}$ and $k_{\text{mir}}$, respectively. 
The matrix in Eq.~\ref{eq:asc_mecc_stiff} is constructed as two identical blocks $2\times2$ as the following.
\begin{equation*}
    \begin{bNiceMatrix} 
        k_{\text{mar}}+k_{\text{mir}} & -k_{\text{mir}}\\
        -k_{\text{mir}} & \;\;\;k_{\text{mir}}\\
    \end{bNiceMatrix}
\end{equation*}
Each block of this type represents the typical mechanical stiffness matrix of a two-stage pendulum, which in our case is the suspension system considered for each optic. Each of these blocks is placed on the "diagonal" of the $4\times4$ matrix. This is because the two suspension systems are mechanically independent in the absence of radiation pressure. In fact, if $K=K_{\textnormal{M}}$, the motion of any stage in one suspension does not lead to the motion of any stage in the other suspension.\\
The matrix $K_{\textnormal{OPT}}$ represents an optically generated stiffness, which we define as: 
\begin{equation}\label{eq:opt_stiff}
    \small
    K_{\textnormal{OPT}} =\frac{2P_{\textnormal{ARM}}}{c}\,\frac{L}{1-g_{\textnormal{ITM}}\;g_{\textnormal{ETM}}}\begin{bNiceMatrix} 
        0 & 0 & 0 & 0\\
        0 & g_{\textnormal{ITM}} & 0 & -1\\
        0 & 0 & 0 & 0\\
        0 & -1  & 0 & g_{\textnormal{ETM}} 
    \end{bNiceMatrix}
\end{equation}
This matrix aims to account for the effect of radiation pressure, as described in~\cite{SIDLES2006167}. To interpret the matrix in Eq.~\ref{eq:opt_stiff}, it should be noted that this matrix is defined as a prefactor $2P_{\textnormal{ARM}}/c$ that multiplies a matrix similar to that in Eq.~\ref{eq:bsm}. The matrix in that equation relates the motion of the optics in a cavity to the beam offset. The matrix in Eq.~\ref{eq:opt_stiff} plays a similar role, connecting the motion of each stage in both suspensions to the beam offset on each test mass. Note that in the way this matrix is structured, a nonzero torque is generated only by the rotation of the mirror stage. This is because only the rotation of the TMs causes the beam offset, which ultimately results in a torque applied to both TMs. All terms in the matrix $K_{\textnormal{OPT}}$ that are related to the motion of the marionette stage are null, since they do not cause the beam offset. Now, it can be more easily understood that the matrix in Eq.~\ref{eq:opt_stiff} is exactly the same matrix as in Eq.~\ref{eq:bsm} but only rewritten in a $4\times4$ form to include the DoFs of the marionette stages. Therefore, the matrix $K_{\textnormal{OPT}}$ is the result of the multiplication of a prefactor that quantifies the amount of radiation pressure force applied to the mirrors and a matrix that defines the beam offset (for a certain rotation of the optics). A force multiplied by an offset in distance is the definition of a torque.\\

ASC loops are designed in a basis where the optical stiffness matrix is diagonal. For each cavity, the eigenvectors of the stiffness matrix define the so-called ``optomechanical modes'' of the cavity. In these eigenvector basis, the stiffness matrix appears diagonal;  more details on the modes of each cavity can be found in Appendix~\ref{appendix:modes}. As the rotation of the optics within a specific cavity results in a beam offset unique to that cavity, each arm cavity is an independent optomechanical system, each with its own optomechanical modes. Despite this independence, the ASC loops use a set of DoFs that combines the modes of each arm cavity. Namely, the following DoFs were controlled for each axis of rotation:
\begin{equation}\label{eq:diag_base}
    \begin{bmatrix} 
        \textnormal{D}_+\\
        \textnormal{C}_+\\
        \textnormal{N}_-\\
        \textnormal{W}_-
    \end{bmatrix} = 
    \begin{bNiceMatrix} 
            t_{1} & \;\;\;t_{2} & -t_{1} & -t_{2}\\
            t_{1} & \;\;\;t_{2} & \;\;\;t_{1} & \;\;\;t_{2}\\
            t_{2} & -t_{1} & \;\; 0 & \;\;0\\
            0 & \;\;\; 0  & \;\;\;t_{2} & -t_{1} 
        \end{bNiceMatrix}
    \begin{bmatrix} 
        \theta_{\textnormal{NI}}\\
        \theta_{\textnormal{NE}}\\
        \theta_{\textnormal{WI}}\\
        \theta_{\textnormal{WE}}
    \end{bmatrix}
\end{equation}
The pairs of values $[t_1\ t_2]$ and $[t_2\ -t_1]$ are eigenvectors of the optical stiffness matrix of each cavity. Since the geometry of both cavities is nearly identical except for slight manufacturing-related differences, these eigenvectors are the same for both cavities. The optomechanical parameters considered are reported in Appendix~\ref{appendix:modes}. For these values, $t_1$ and $t_2$ were calculated as $0.64$ and $0.76$, respectively. The selection of the DoFs in Eq.~\ref{eq:diag_base} is motivated by optical considerations, where each produces a unique effect on the beam~\cite{casanueva2017}; see Fig.~\ref{fig:asc_dofs}.\\

The ultimate goal of the modeling is to evaluate the effectiveness of the ASC system by quantifying the degree to which angular motion, within the detection frequency band ($f > 10$Hz), contributes to the detector output. To achieve this, the mechanics, as previously described, and the control scheme used during O3 have been incorporated into \finesse~model of the Virgo detector~\cite{VirgoKat, FINESSE, finesse3}. Additional details on the optomechanical parameters considered can be found in Appendix~\ref{appendix:modes}. The magnitude of the Open Loop Transfer Function (OLTF) for each loop is depicted in Fig.~\ref{fig:oltf}, while the control system architecture is outlined in Appendix~\ref{appendix:control}. With a specific emphasis on mechanics, only the ``pure'' mechanics have been imported into the model, with a stiffness matrix given by $K=K_{\textnormal{M}}$. This is because \finesse~includes the effects of radiation pressure on the mechanics by default, eliminating the need to add an optical stiffness to the imported mechanics. For more details on how \finesse~handles radiation pressure, refer to~\cite{Brown2016}. The analytical model of radiation pressure proposed here was used to verify the result of \finesse, which proved to be a suitable tool for the modeling that we intended to perform. In fact, as shown in the Appendix~\ref{appendix:modes}, the results of the two models, analytical and \finesse, are perfectly aligned.\\
Beyond validation, our analytical approach facilitates a straightforward and effective calculation of the frequency response of the optomechanical system, without the need to go through a complete interferometer model. To do this, we convert Eq.~\ref{eq:asc_motion} into the Laplace domain as follows:
\begin{equation}
-s^2 I \theta(s) - K \theta(s) = T(s)
\end{equation}
The transfer matrix that links a torque to the resulting rotation of the DoFs of the mechanics can simply be calculated as:
\begin{equation}\label{eq:asc_plant}
\frac{\theta}{T} = \left(-s^2\mathbb{I} - I^{-1} K\right)^{-1} I^{-1}
\end{equation}

\section{External disturbances}\label{sec:asc_3}

Accurate evaluation of control noise involves considering external disturbances that influence optic movements, according to the dynamics defined by the control system and mechanics. Through the analysis of experimental data, we identified and characterized various external disturbances. In particular, these include sensor noise, seismic noise, and a disturbance that we refer to as ``longitudinal noise''. This term denotes an additional torque applied to the marionette due to imbalances in the actuators used for longitudinal control of the detector. The actuators connected to the marionette play a role in the length control system. When these actuators experience imbalances, the force applied for length control generates an additional torque. In Appendix~\ref{appendix:noise}, we provide a more detailed description of all the noises that have been considered. In the appendix, we show that by propagating these considered noises through the modeled loops, we can effectively reconstruct the error signals obtained experimentally. Therefore, the noise set considered is a complete set to reconstruct the residual movement of the TMs.

\begin{figure}[!b]
	\centering
	\includegraphics[width=\linewidth]{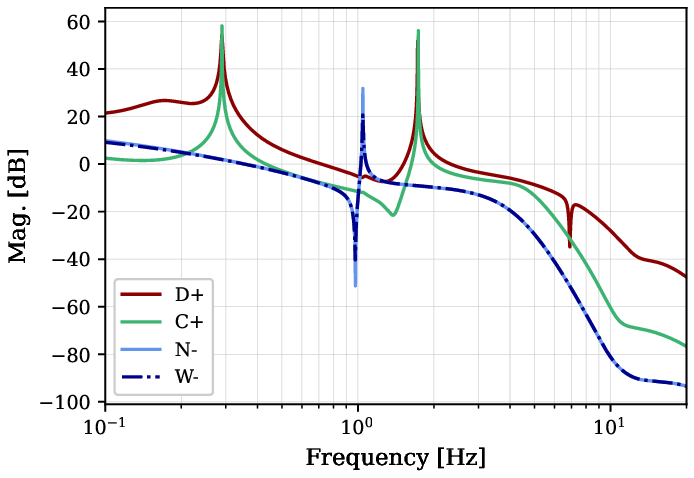}
	\caption{\label{fig:oltf} Magnitude of the open-loop transfer function of each loop.}
\end{figure}

From the propagation of noises in the model, we calculated the residual movement of the TMs, which is shown in Appendix~\ref{appendix:noise}. From the movement of the TMs, it was possible to reconstruct the BSM using Eq.~\ref{eq:bsm}. The result is shown in Fig.~\ref{fig:projections_tot}, which shows a measurement of the BMS on one of the TMs and compares it with the model output. The contribution of each external disturbance to the BSM is shown in the Appendix~\ref{appendix:noise}. It should be noted that this is the first time the measured BSM has been accurately reproduced using a numerical model of the complete optomechanical system.

\section{Noise projections to the detector output}\label{sec:asc_4}
\begin{figure}[!t]
	\centering
	\includegraphics[width=\linewidth]{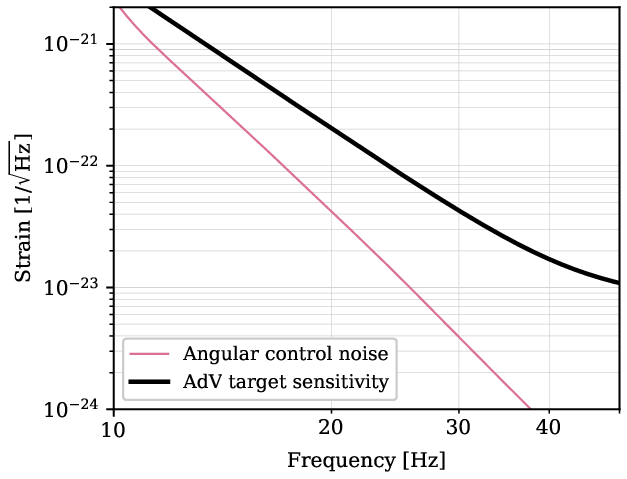}
	\caption{\label{fig:projections_tot650}The figure shows the level of angular control noise calculated with the model assuming AdV were operated at full power. Even in this case, the detector would not be constrained by this noise. Specifically, this noise projection was calculated based on a circulating power in the cavities of $650$~kW, approximately five times higher than that used during O3. The target sensitivity for AdV during O3 is represented in black.}
\end{figure}
Continuing to follow the guidelines outlined starting from Eq.~\ref{eq:angle2length}, it should be clear that once the movement of the TMs and the BSM is calculated, it is possible to compute the amount of angular control noise coupled to the detector output. However, as previously mentioned, the complexity increases when considering the entire detector, compared to the scenario described by Eq.~\ref{eq:angle2length}. The angular motion of the TMs coupled to the detector output was calculated using \finesse, and the corresponding result is shown in Fig.~\ref{fig:projections_tot}. More comprehensive details on the methodology used can be found in Appendix~\ref{appendix:proj}. In the appendix, we show the contribution to angular control noise from each controlled DoF. Fig.~\ref{fig:projections_tot} shows the total level of control noise.\\
The noise projection generated by the model confirms that Advanced Virgo was not limited by the angular control noise during O3. The model output shows excellent agreement with experimental measurements, indicating the success of the methodology developed for evaluating the residual rotation of the TMs and its subsequent impact on the detector output.This establishes a robust foundation for modeling future detector configurations and future generation detectors.\\

The noise projection was repeated for the case where Virgo was operated at full power. The result is shown in Fig.~\ref{fig:projections_tot650}, indicating that Virgo would not be limited by angular noise even in this scenario. To produce this projection, the input noise and the architecture of the control system were left unchanged. The key change resulting from the increase in power was the impact of the radiation pressure on the mechanical components. The control filters were adapted to the new plant, but they were still designed to ensure a match with the requirements for the residual motion of the TMs.

\subsection{Implications for the Einstein Telescope}
\begin{figure}[!b]
	\centering
	\includegraphics[width=\linewidth]{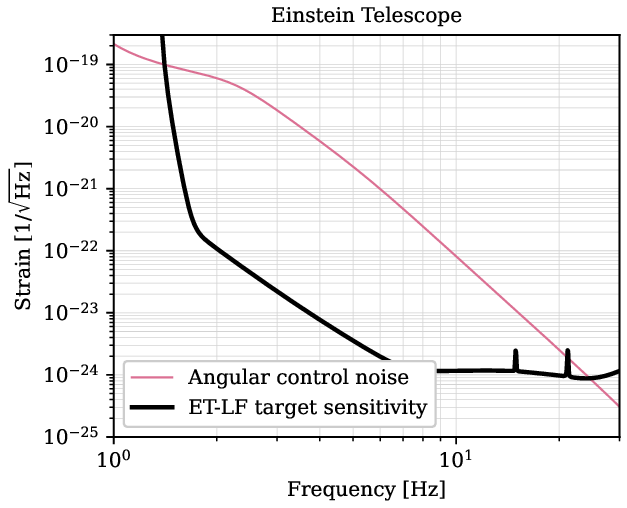}
	\caption{\label{fig:proj_et}Projection of angular noise on the low-frequency sensitivity of ET.}
\end{figure}
We are now in a position to extend the application of the ASC model to the Einstein Telescope case. Fig.~\ref{fig:proj_et} illustrates a prediction of the angular noise level in ET. It is evident that this noise has the potential to pose a significant challenge to ET. However, it is essential to emphasize that the true result here is the ability to track the noise level and that the projection itself is likely an exaggerated estimate.\\
This projection was obtained by directly transferring the AdV ASC system to ET, assuming that almost everything remains unchanged, which is likely not the case. Specifically, we applied the control architecture described in Section~\ref{sec:asc_4} to the ET segment designed for low-frequency measurements, known as ET-Low-Frequency (ET-LF). The optical layout considered is the one implemented in the corresponding \finesse~model~\cite{EtKat}. As for the mechanics, we considered a suspension inspired by the Virgo superattenuator; more details are provided in Appendix~\ref{appendix:ETmech}.
The two elements that require more attention to understand the noise projection are the control filters and external disturbances. Regarding the control filters, we employed those from AdV and adapted them to the new system. Since the requirements for the residual rotation of the TMs in ET have not yet been defined, the filters have been designed to roughly match the requirements set for Virgo. However, well-defined requirements for ET are essential to determine the loops gain. Without established requirements, the design of the control filters becomes somewhat arbitrary. In simple terms, for the same input noise, a system that requires more extensive control will generate a higher amount of control noise. An example of this effect can be observed in Fig.~\ref{fig:oltf}, where we compare the OLTF of D$_+$ with that of C$_+$. Both modes are hard modes and, therefore, share the same plant. As explained in the previous Section~\ref{sec:asc_2}, from a purely mechanical perspective, the arm cavities are identical systems, each with its hard and soft modes. If the TMs in each cavity move in a D$_+$  or C$_+$ manner, there is no mechanical distinction between them. As a result, it might be logical to assume that the same controller could be employed for both DoFs, and thus the OLTF should be identical. However, this is clearly not the case. As explained in Section~\ref{sec:asc_2}, while the common or differential motion of the TMs may be identical from a mechanical point of view, this is not true from an optical point of view. The motion of D$_+$ defines the differential tilt between the arm cavities, affecting the quality of the dark fringe in the output port. Simply put, the less D$_+$ moves, the better the quality of the dark fringe. Consequently, the control loop for this specific DoF requires a higher gain over a wider frequency range, resulting in a different filter design. This filter starts to roll off later and with a gentler slope compared to the filter for C$_+$. As a result, D$_+$ becomes the main source of injected angular control noise at the detector output - see Appendix~\ref{appendix:proj}. This happens despite the fact that D$_+$ has a high-frequency sensing noise almost an order of magnitude lower than that of C$_+$ - see Appendix~\ref{appendix:noise}.\\
The other elements that require discussion are external disturbances. The noises considered are identical to those discussed in Section~\ref{sec:asc_4}. For almost all these noises, there is undoubtedly room for hardware improvement to reduce the noise level.\\

There is one last difference worth highlighting between the noise projection for Virgo and that for ET. For the Einstein Telescope case, the contribution of the soft modes to the projection (N$-$ and W$-$) was excluded. In the scheme inherited from Virgo, these degrees of freedom are controlled using optical levers, known to be extremely noisy sensors; see Appendix~\ref{appendix:noise}. The projection includes only the contributions of the hard modes (D$+$ and C$+$). These degrees of freedom are controlled using QPDs. Notably, these sensors exhibit significantly better noise performance compared to optical levers. The choice to exclude the soft modes from the projection was made to avoid any potential misconception that ET might be limited by angular control noise solely due to the potential use of optical levers. Instead, to accurately assess whether ET will be limited by angular control noise or not, a thorough analysis involving the entire control system is required. This analysis must be based on well-established requirements for the residual motion of various DoFs. These requirements will then inevitably be translated into requirements for the mechanics and external disturbances.


\section{Conclusions}\label{sec:asc_5}

In this study, we have successfully developed a model of the Advanced Virgo angular control system, confirming that it has not been limited by angular control noise, despite facing external disturbances comparable to those of aLIGO. The construction of the model has led to a thorough characterization of the angular control system, which has shown that to model the angular control noise accurately, a comprehensive understanding of the various external disturbances is essential. Furthermore, using the developed and validated model, we have shown that Virgo would not be limited by angular control noise even when operated at full power. More generally, the remarkable agreement between our ACS model and experimental data represents a significant milestone, providing a solid foundation for modeling the low-frequency design of future-generation detectors.\\
Finally, we have shown that angular control noise has the potential to obstruct ET goals. We have explained how formulating control requirements is crucial for accurately assessing whether the ET will be limited by angular control noise or not.

\newpage
\appendix
\section{\label{appendix:modes}Additional details of the mechanics}

ASC loops are designed on a base where the plant, or equivalently the optical stiffness matrix, is diagonal. The diagonalized form of $K_{\textnormal{OPT}}$ can be written as:
\begin{equation}
    K_{\textnormal{OPT,\,D}} = S^{-1}\, K_{\textnormal{OPT}} \; S
\end{equation}
Each column of $S$, the diagonalizing matrix, corresponds to an eigenvector of the matrix $K_{\textnormal{OPT}}$. For each arm cavity, the transformation to the basis in which $K_{\textnormal{OPT}}$ is diagonal can be represented as follows:

\begin{equation}\label{eq:hard_soft}
	[\theta_+ \; \theta_-]^{T} = S \cdot [\theta_{\textnormal{ITM}} \; \theta_{\textnormal{ETM}}]^{T}
\end{equation}
For a more detailed analytical treatment of the eigenvectors of $K_{\textnormal{OPT}}$, please refer to \cite{Barsotti_2010}. Regarding our discussion, the important information is that the eigenvectors are a pair of vectors of the form $[t_1\ t_2]$ and $[t_2\ -t_1]$. Therefore, the change of basis matrix $S$ can be written as:
\begin{equation}\label{eq:S}
S =
\begin{bmatrix} 
t_{1} & t_{2}\\
t_{2} & -t_{1}\\
\end{bmatrix} 
\end{equation}
The basis in which the optical stiffness matrix appears in a diagonal form is commonly known as the basis for the ``hard and soft modes''.  When the mirrors move in the same direction, the beam tilts, resulting in a radiation pressure torque that tends to bring the optics back to their equilibrium position. This additional restoring effect strengthens the mechanical stiffness, which is why $\theta_{+}$ is called the hard mode. Conversely, when the mirrors move in opposite directions, the beam shifts, and the resulting radiation pressure effect creates a torque that reinforces the displacement of the optics away from their equilibrium position. This additional anti-restoring effect softens the mechanical stiffness, which is why $\theta_{-}$ is known as the soft mode.

\begin{figure}[!t]
	\centering
	\includegraphics[width=\linewidth]{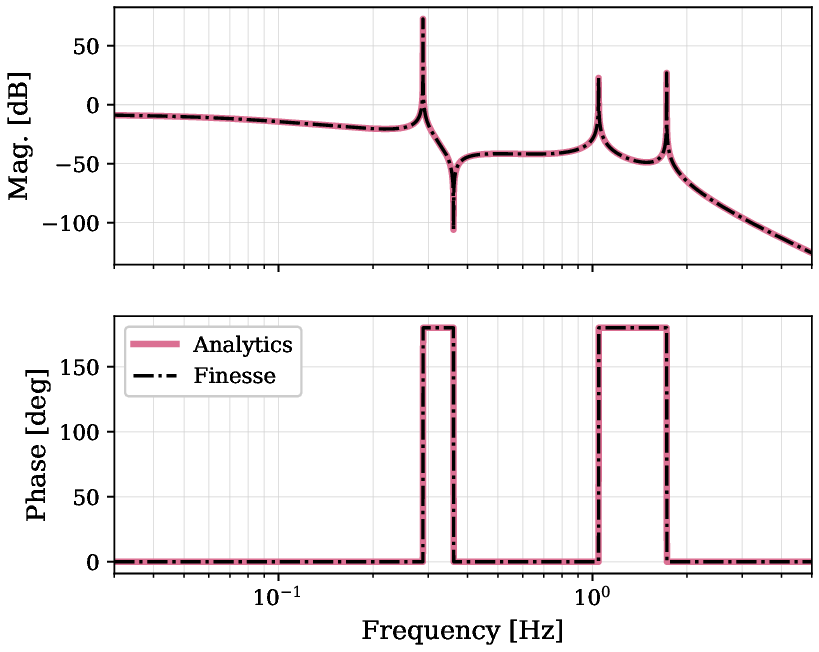}
	\caption{\label{fig:plant_finesse}Transfer function connecting the torque applied to the marionette of NI suspension to the motion of NE, calculated both through our analytical model and using~\finesse. The results obtained with the two methods are well aligned, confirming that Finesse is a suitable tool for the modeling we performed. Furthermore, the agreement between the two results demonstrates that the analytical model developed is a highly effective and straightforward tool that allows the inclusion of the radiation pressure effect on the TMs suspension.}
\end{figure}
\begin{table}[!b]
  \caption{Mechanical parameters of the suspension.}
  \vspace{0.1cm}
  \label{tab:asc_mechanics}
  \begin{ruledtabular}
    \begin{tabular}{lc}
      \textbf{Parameter} & \textbf{Value} \\
      \hline
      Marionette Stiffness [Nm/rad] & 0.07 \\
      Test Mass Stiffness [Nm/rad] & 18.7 \\
      Marionette Inertia [kg·m$^2$] & 3.65 \\
      Test Mass Inertia [kg·m$^2$] & 0.46 \\
    \end{tabular}
  \end{ruledtabular}
\end{table}
\begin{figure}[!t]
	\centering
	\includegraphics[width=\linewidth]{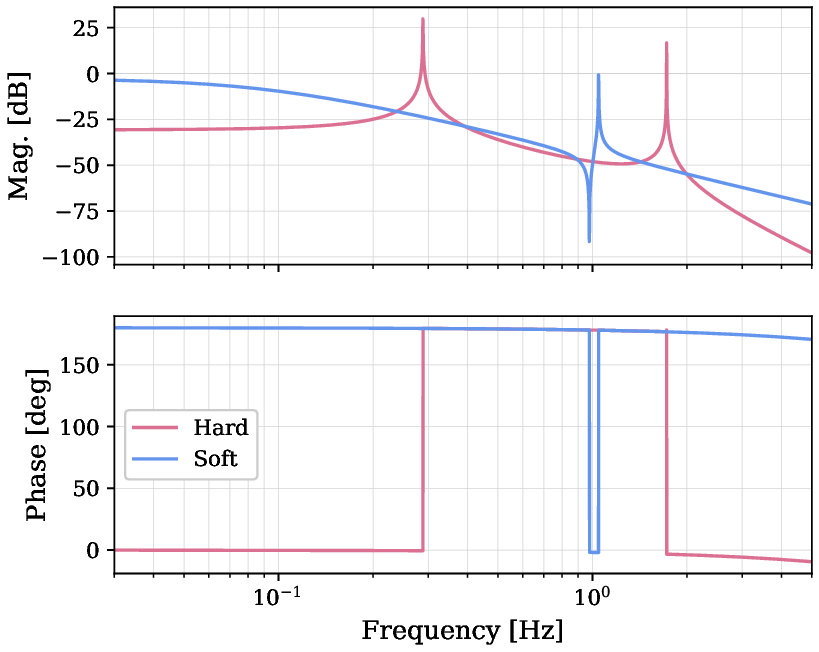}
	\caption{\label{fig:plant_hs} The torque to rotation transfer function of the optomechanical plant on the basis of hard and soft modes. For both curves, the
torque is applied to the marionette, while the resulting motion
is measured at different levels. For the hard mode, the output
motion is read at the test mass level, whereas for the soft mode, it is measured at the marionette level.}
\end{figure}
\begin{figure*}[!t]
	\centering
	\includegraphics[width=0.85\textwidth]{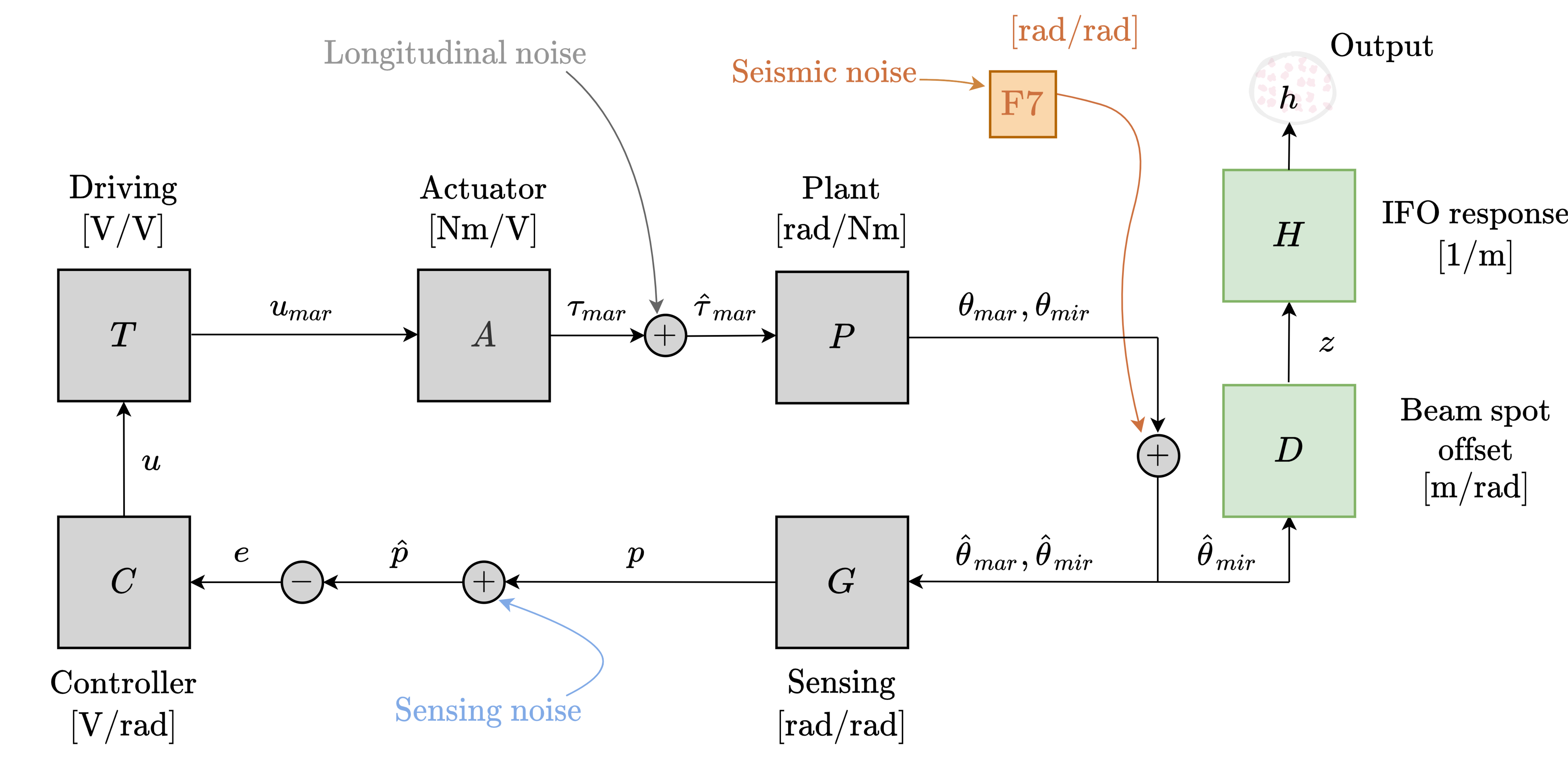}
	\caption{\label{fig:block_diagram} The gray blocks represent the block diagram of Advanced Virgo ASC loops. The diagram shows the entry points of external disturbances into the loops, a description of which can be found in Section~\ref{sec:asc_3}. The F7 block connects the residual motion of the suspension stage above the marionette, known as Filter 7, to a torque applied to the marionette. The green branch of the diagram illustrates how the mirror motion couples to the output of the detector; its elements are explained in detail in Appendix~\ref{appendix:proj}.}
\end{figure*}
For the optical parameters considered in our modeling, the values of $t_1$ and $t_2$ have been calculated as $0.64$ and $0.76$, respectively.  Specifically, the circulating power in the arms was considered to be $131.5$~kW.  Regarding the RoC values and lengths that have been taken into account, we use the specifications provided in the \finesse~model of Virgo~\cite{VirgoKat}. In terms of RoCs and lengths, the two arm cavities are almost identical, with only small manufacturing discrepancies. For this reason, the values of $t_1$ and $t_2$ turned out to be practically the same for both cavities. Regarding the mechanical parameters of each TM suspension, these details are provided in Tab.~\ref{tab:asc_mechanics}. All optomechanical parameters were used to calculate the torque-to-rotation transfer functions through the analytical model shown in Section~\ref{sec:asc_2}, specifically using Eq.~\ref{eq:asc_plant}; an example of the result is shown in Fig.~\ref{fig:plant_finesse}. Lastly, in Fig.~\ref{fig:plant_hs}, the torque-to-rotation transfer functions are shown at the base of the hard and soft modes. These have been calculated as follows:
\begin{equation}
   P_{\textnormal{D}} = S^{-1}\, P \; S
\end{equation}
Where $P$ denotes what we will refer to in Appendix~\ref{appendix:control} as the plant and $P_{\textnormal{D}}$ is its diagonal form. $P$ is connects the torque applied to the suspension to the rotation of the TMs and marionettes. In essence, it is a transfer matrix, and each element is a transfer function of the type shown in Fig.~\ref{fig:plant_finesse}. We focus solely on elements connecting the torque applied to the suspension to the motion of other DoFs, not those linking a torque applied to the TMs to DoF motion. This choice is due to the control scheme in O3, which exclusively employed actuators attached to the suspension (see Appendix~\ref{appendix:control}). Nonetheless, Eq.~\ref{eq:asc_plant} (or equivalently~\finesse) facilitates the calculation of these terms.

\section{\label{appendix:control}The control system architecture}
 In this section, we will illustrate the control system architecture following the structure of the block diagram in Fig.~\ref{fig:block_diagram}.\\
The feedback controller ${C\in\mathcal{R}}^{4\times4}$, uses an error signal:
\begin{equation}
	e = [e_{\textnormal{D}_+} \; e_{\textnormal{C}_+} \; e_{\textnormal{N}_-} \; e_{\textnormal{W}_-}]^{T} 
\end{equation}
\begin{table}[b]
\caption{Sensing and driving matrices are in units [rad/rad]. The sensing matrix, shown on the left, was obtained from the analysis of experimental data and then imported into the \texttt{\finesse} model. The driving matrix is shown on the right.}
\label{tab:asc_driving}
\vspace{0.1cm}
\small 
\centering
\begin{minipage}{0.45\linewidth}
\centering
\begin{tabular}{|c|c|c|c|c|}
\hline
& \textbf{NI} & \textbf{NE} & \textbf{WI} & \textbf{WE} \\
\hline
\textbf{D$_+$} & -1.73 & -2.13 & 1.95 & 2.14 \\
\hline
\textbf{C$_+$} & 5.37 & 6.12 & -1.93 & -2.16 \\
\hline
\textbf{N-} & -1.00 & 1.00 & 0 & 0 \\
\hline
\textbf{W-} & 0 & 0 & -1.00 & 1.00 \\
\hline
\end{tabular}
\end{minipage}%
\hspace{2em} 
\begin{minipage}{0.45\linewidth}
\centering
\begin{tabular}{|c|c|c|c|c|}
\hline
& \textbf{D$_+$} & \textbf{C$_+$} & \textbf{N-} & \textbf{W-} \\
\hline
\textbf{NI} & 0 & 0.50 & -0.52 & 0 \\
\hline
\textbf{NE} & 0 & 0.50 & 0.48 & 0 \\
\hline
\textbf{WI} & 0.50 & 0.50 & 0 & -0.54 \\
\hline
\textbf{WE} & 0.50 & 0.50 & 0 & 0.46 \\
\hline
\end{tabular}
\end{minipage}
\end{table}
to produce a control signal ${u\in\mathbb{R}}^{4}$ for each DoF. The driving matrix ${T\in\mathbb{R}}^{4\times4}$ applies a coordinate transformation to the actuation signal to obtain the actuation to the marionette of each mirror:
\begin{equation}
	u_{\textnormal{mar}} = [u_{\textnormal{NI}} \; u_{\textnormal{NE}} \; u_{\textnormal{WI}} \; u_{\textnormal{WE}}]^{T} 
\end{equation}
The actuation to each marionette is converted into a torque by the matrix ${A\in\mathcal{R}}^{4\times4}$, which contains the dynamic of the actuators. The optomechanical plant $P$, which has been defined in the previous section, connects the torque applied to the marionette ${\tau_{\text{mar}}\in\mathbb{R}}^{4}$ to the rotation of the two suspension stages ${\theta_{\textnormal{mar}}},{\theta_{\textnormal{mir}}\in\mathbb{R}}^{4}$, the marionette, and the mirror, respectively. The sensing matrix ${G\in\mathbb{R}}^{4\times4}$ connects the rotation of the optics and that of the marionette to the error signal of each DoF, which is denoted by:
\begin{equation}
	p = [p_{\textnormal{D}+} \; p_{\textnormal{C}+} \; p_{\textnormal{N}-} \; p_{\textnormal{W}-}]^{T} 
\end{equation}
\begin{figure*}[!t]
    \centering
        \includegraphics[width=\textwidth]{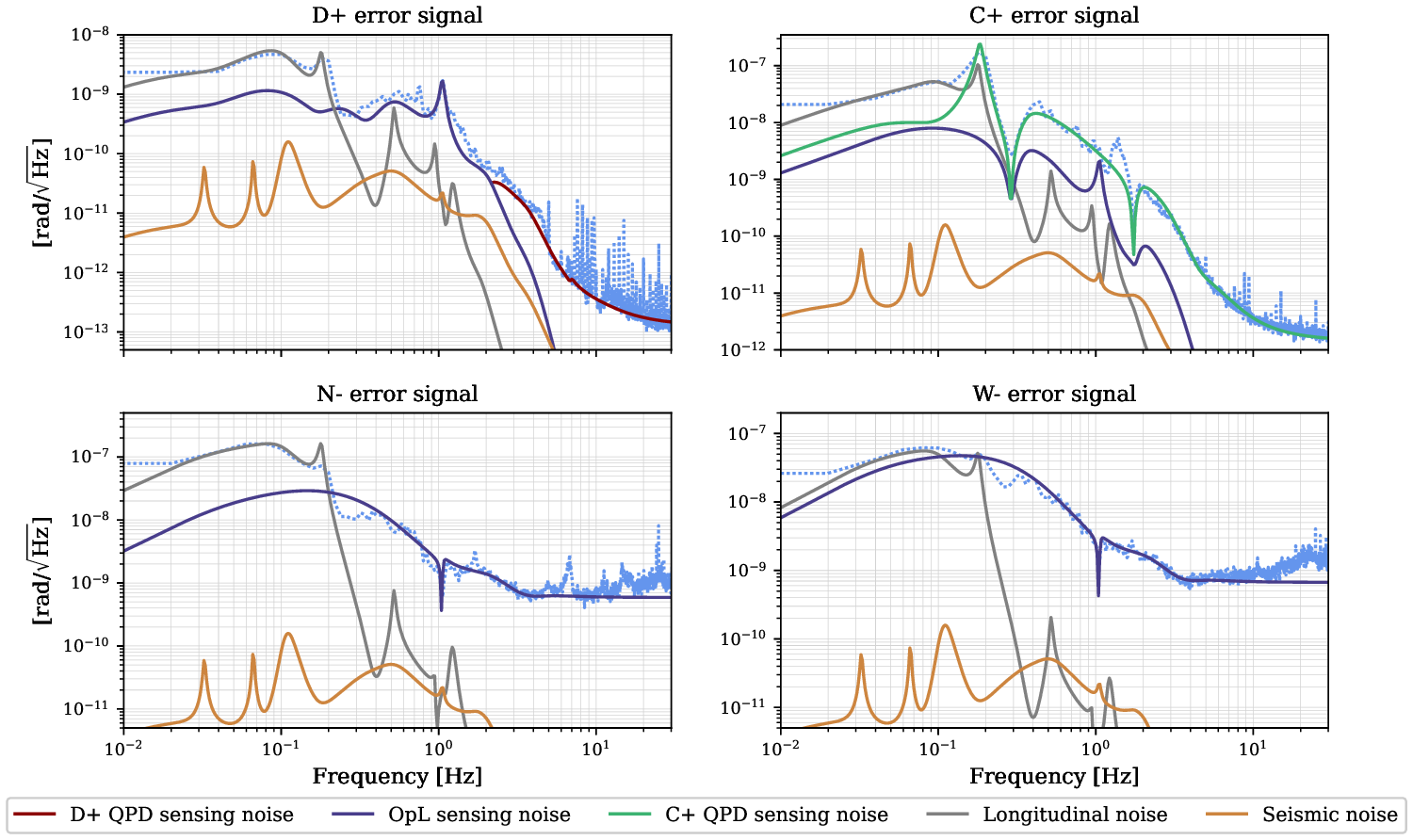}
        \caption{\label{fig:errors}Noise budget of the ASC error signals. The contribution to each error signal was calculated using the \finesse~model, with the methodology explained at the beginning of the Appendix~\ref{appendix:noise}.}
\end{figure*}
Considering the marionette movement is necessary because the O3 sensing scheme utilised optical levers reading at the marionette level to obtain $p_{\textnormal{N}_-}$ and $p_{\textnormal{W}_-}$. In the case of D$_+$ and C$_+$, QPDs were used as sensors. Specifically, Ward's technique~\cite{Morrison:94} was employed for D$_+$, while for $p_{\textnormal{C}_+}$, the beam's DC position on the dedicated photo detector was measured. A more detailed description of the angular sensing scheme used during O3 can be found in~\cite{galaxies8040085}.\\
For the specifications of the individual components of the loop, the sensing and driving matrices can be found in Tab.~\ref{tab:asc_driving}. The control filters used are the same as those used during O3.\\

We conclude this section with a brief discussion of the sensing and driving matrices in Tab.~\ref{tab:asc_driving}. The sensing matrix transforms the coordinate system from that of the individual optics' movements to that of the optomechanical modes. On the contrary, the driving matrix performs the reverse transformation, shifting from the coordinate system of the optomechanical modes to that of the individual optics. At this point, it should be easy to understand that if the sensing matrix coincides with the matrix $S$ in Eq.~\ref{eq:S} and the driving matrix with $S^{-1}$, then the plant is perfectly diagonal. However, when examining the sensing matrix, it is evident that it does not coincide with the matrix $S$, leading to couplings between the DoFs. The driving matrix has been experimentally tuned to mitigate these couplings, specifically to limit the coupling shown in Fig.~\ref{fig:opl2dp}; see Appendix~\ref{appendix:noise}.

\section{External disturbances}\label{appendix:noise}
From the analysis of the experimental data, we identified and characterized a series of external disturbances. These noises were then propagated through the modeled loops to calculate the residual movement of the optics. With this we mean that using \finesse, the transfer function was calculated from the noise input point to the error signal point. Then, this transfer function was multiplied by the input noise to calculate the contribution to the error signal produced by that specific noise. In Fig.~\ref{fig:block_diagram}, we have marked the input point for each external noise source that we considered. Fig.~\ref{fig:errors} illustrates the resulting error signals for each input noise. The figure highlights that the input noises considered are adequate to reconstruct all error signals. Therefore, the set of noises serves as a comprehensive description of the inputs of the system.  

\begin{figure}[!t]
	\centering
	\includegraphics[width=\linewidth]{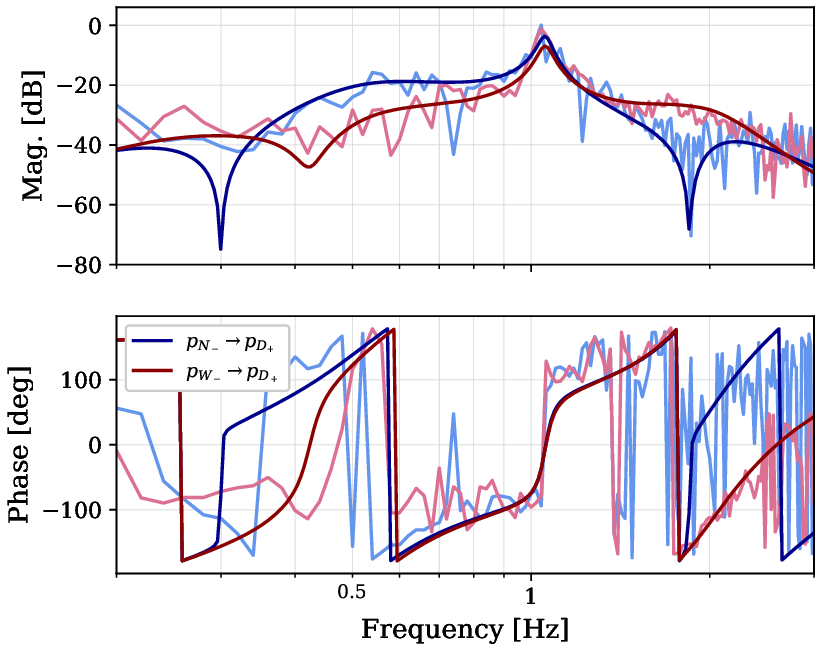}
	\caption{\label{fig:opl2dp}A direct comparison of the model's output, a transfer function, with its experimental counterpart. The figure shows the transfer functions that map the error signal of the soft modes to that of D$_{+}$ is shown below. The light-colored curve represents the measurement, while the dark-colored curve represents the model output.}
\end{figure}

More generally, the output of the \finesse~model is a transfer function from any point on the diagram in Fig.~\ref{fig:block_diagram} to any other point on the same diagram. Therefore, the modeling methodology just outlined to calculate error signals can be extended to calculate any signal of interest. It involves calculating the transfer function from the noise input point to the desired signal and then multiplying that transfer function by the input noise.\\
A comparison between the transfer functions calculated from the experimental data and those calculated through the model is shown in Fig.~\ref{fig:opl2dp}.\\

Regarding the external disturbances that we have considered, it is important to note that some of them are influenced by environmental conditions. For our analysis, we selected a data set in which the environmental conditions led to a relatively large movement of the mirrors and, therefore, to a relatively large angle-to-length coupling. We aimed to confirm our model's precision under extreme working conditions and also helped in isolating and identifying the individual input contributions.\\
We will now proceed to describe each of the noises considered.

\begin{figure}[!b]
	\centering
	\includegraphics[width=\linewidth]{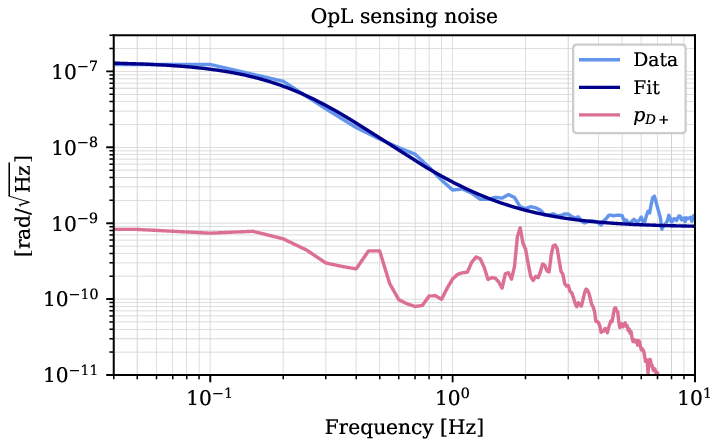}
	\caption{\label{fig:opl_noise}The red measurement indicates the D$_{+}$ error signal while the light blue measurement represents the same DoF measured using OpLs at the marionette level, and its corresponding fit is shown in dark blue.}
\end{figure}

\paragraph{Optical levers sensing noise}
In Fig.~\ref{fig:opl_noise}, the red curve indicates the measurement of $p_{\textnormal{D}_+}$, while the blue curve represents the motion of the same degree of freedom as measured by OpLs. Due to the sensor hitting its noise floor, the latter signal is higher than $p_{\textnormal{D}_+}$ in the whole frequency range. The dark blue curve in the plot corresponds to the measurement fit, representing the level of OpLs sensing noise that we have taken into account.\\

OpLs are used to control soft modes, specifically N$_-$ and W$_-$. The noise from these sensors is then re-introduced into the control loops, leading to motion in these modes. However, due to the non-diagonal nature of the system, this noise also induces motion in D$_{+}$, as shown in Fig.~\ref{fig:errors}. The transfer function, shown in Fig.~\ref{fig:opl2dp}, is responsible for this coupling.

\paragraph{C$_{+}$ QPD sensing noise}

\begin{figure}[!b]
	\centering
	\includegraphics[width=\linewidth]{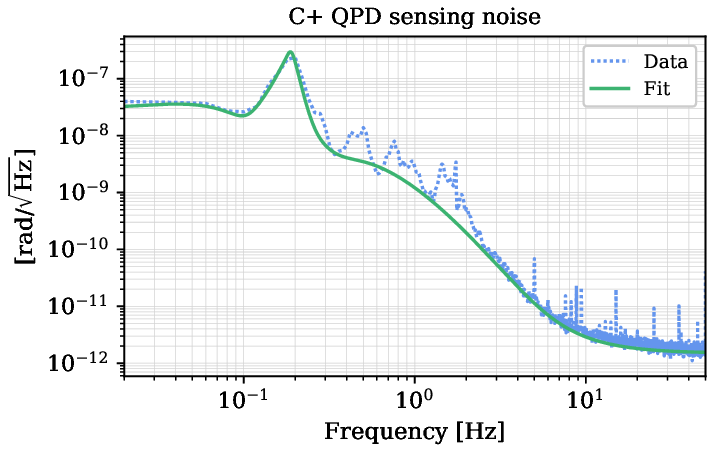}
	\caption{\label{fig:cp_noise}A measurement of $p_{\textnormal{C}_+}$ is shown along with its fit in green, after being multiplied by the inverse of the loop suppression.}
\end{figure}
In Fig.~\ref{fig:errors} is a measurement of the C$_{+}$ error signal. The presence of two pronounced troughs at the mechanical frequencies, around $300$~ mHz and $2$~ Hz, respectively, suggests that this signal was dominated by the sensor noise. The feedback action of a loop suppresses a signal by a factor of one minus the open-loop gain. If the signal consists of noise and the loop gains at the mechanical frequencies, the result are these deep troughs in the sensor noise. Fig.~\ref{fig:cp_noise} shows a measurement of $p_{\textnormal{C}_+}$ that has been divided by the loop suppression, resulting in an equivalent open-loop noise measurement. 
The structure at $\sim200$~ mHz in the $p_{\textnormal{C}_+}$ signal requires separate discussion. This was found to be coherent with the longitudinal motion of the lens that served as a placeholder for the signal recycling mirror. The movement of that lens resulted in jitters in the beam of the C$_{+}$ QPD, which ultimately produced the $p_{\textnormal{C}_+}$ signal. Although this is not strictly sensing noise, this signal is reinjected into the C$_{+}$ loop as much as the sensor noise. This is the reason why we have included it as part of the C$_{+}$ QPD sensing noise.

\begin{figure}[!t]
	\centering
	\includegraphics[width=\linewidth]{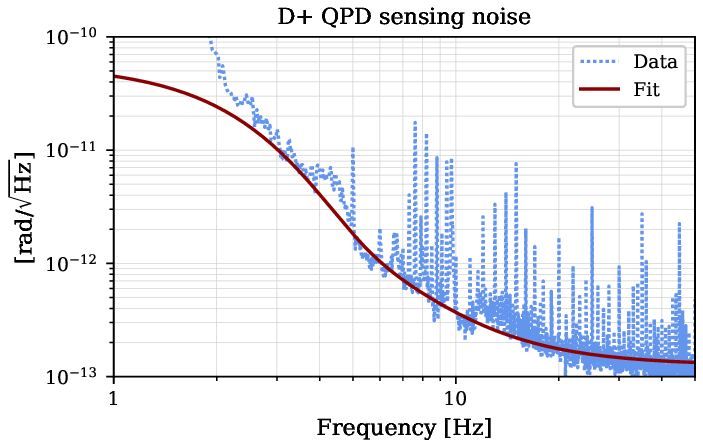}
	\caption{\label{fig:dp_noise}The measurement of $p_{\textnormal{D}_+}$ is displayed along with its fit in red, following the division by the loop suppression.}
\end{figure}
\begin{figure}[!t]
	\centering
	\includegraphics[width=\linewidth]{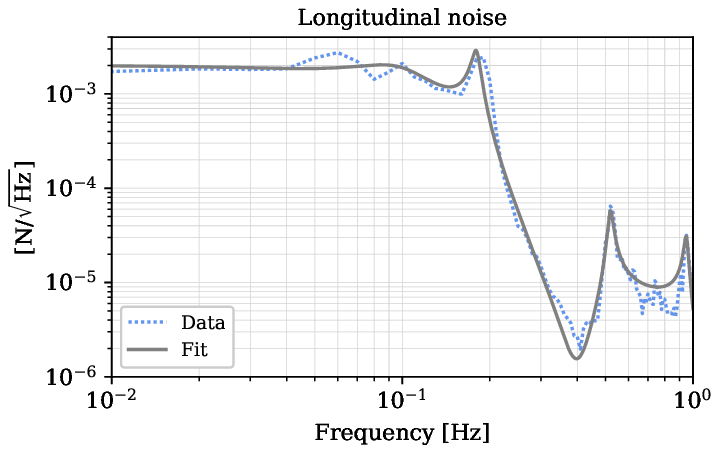}
	\caption{\label{fig:z_corr}The measurement of the force exerted on the marionette for length control is depicted along with its fit in grey.}
\end{figure}

\paragraph{D$_{+}$ QPD sensing noise}
As shown in Fig.~\ref{fig:errors}, the D$_{+}$ sensor detects movements resulting from various sources of noise up to $\sim2$~Hz. The sensing noise of D$_{+}$ QPD has been regarded as anything beyond that frequency. Similarly to what was done for C$_{+}$ sensor noise, the measurement of $p_{\textnormal{D}_+}$ shown in Fig.~\ref{fig:dp_noise} has been divided by loop suppression to remove the influence of the loop. The red curve represents the fit of the measurement for the frequency range of interest. This curve corresponds to the level of D$_{+}$ QPD sensing noise that we have considered.
\paragraph{Longitudinal noise}
The actuators attached to the marionette were used as part of the length control system to correct for low-frequency disturbances. In the event that these actuators are imbalanced, the force applied for length control $F_{z}$ will generate an additional torque on the marionette as:
\begin{equation}
    \eta_{z} = \rho r F_{z}
\end{equation}
\begin{table}[!b]
    \caption{Measured imbalance levels of actuators and their corresponding values from the model. The actuators are located at a distance of $r=0.3$~m from the rotational axis. The length driving scheme exclusively employs actuators for the end test masses.}
    \vspace{0.2cm}
    \label{tab:z_corr_tab}
    \centering
    \small 
    \begin{tabular}{|l|r|r|}
        \hline
        \textbf{Imbalance $\rho$} & \textbf{Model} & \textbf{Measured} \\
        \hline
        \textbf{NE} & $0.21\%$ & $0.10\%$ \\
        \hline
        \textbf{WE} & $0.04\%$ & $0.03\%$ \\
        \hline
    \end{tabular}
\end{table}
Here, $r$ denotes the position of an actuator relative to the rotation axis, and $\rho$ is a scaling factor that represents the degree of imbalance between all actuators attached to a marionette. The force applied to a marionette for length control is depicted in blue in Fig.~\ref{fig:z_corr}, along with its fit in gray. In particular, coherence was observed between this signal and the angular error signals in the low-frequency range ($f<200$~mHz), indicating that the rotation of the mirrors was driven by the length correction. Longitudinal noise has been defined as the signal in Fig.~\ref{fig:z_corr} multiplied by $\rho r$. The result is a torque that has been applied to each marionette involved in the length control scheme. The level of imbalance of the actuators of each marionette was adjusted to effectively replicate the error signals within the coherent frequency range using the model. As shown in Tab.~\ref{tab:z_corr_tab}, the degree of imbalance obtained from the model is in close agreement with the measured value.

\begin{figure}[!b]
	\centering
	\includegraphics[width=\linewidth]{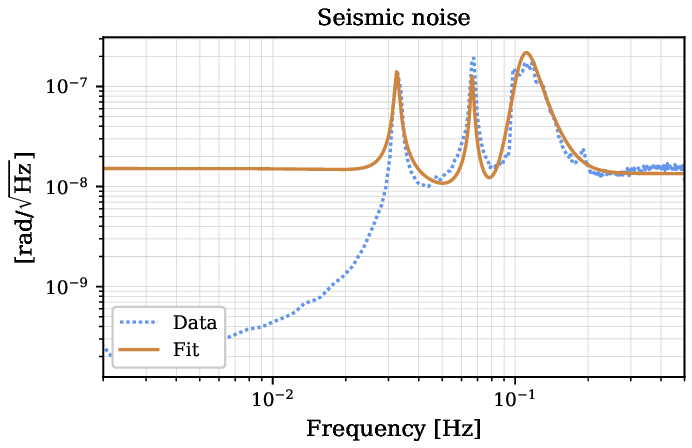}
	\caption{\label{fig:seism_noise}The measurement of a local sensor used to monitor the motion of Filter 7. The brown curve, labeled as ``fit'' for consistency with other plots, represents the level of seismic noise we have considered}
\end{figure}

\paragraph{Seismic noise}
We considered the movement of the suspension point of the marionette as seismic noise, specifically the motion of Filter 7. This is the suspension stage positioned directly above the marionette. The motion of the ground is transmitted through the suspension, causing the movement of Filter 7. This, in turn, induces motion in the marionette, thereby entering the ASC loops.\\
Fig.~\ref{fig:seism_noise} illustrates the measurement obtained from a local sensor responsible for monitoring the motion of Filter 7. The signal captured by this sensor consists mainly of noise, except for three distinct structures occurring between $30$ and $120$~mHz. The curved segment of the signal, up to around $30$~mHz, is a result of the sensor noise being ``bent'' by the influence of the local loop, rather than representing the actual motion of Filter 7. In more general terms,  the local sensor does not measure for the most part the actual motion of Filter 7; instead, it captures mainly noise. The brown curve represents the level of seismic noise that we have taken into consideration. It is calibrated based on the local sensor noise, which, as we have explained, surpasses the actual motion of Filter 7. Hence, this assessed noise level should be regarded as an upper limit.\\

\begin{figure}[!t] 
	\centering
	\includegraphics[width=\linewidth]{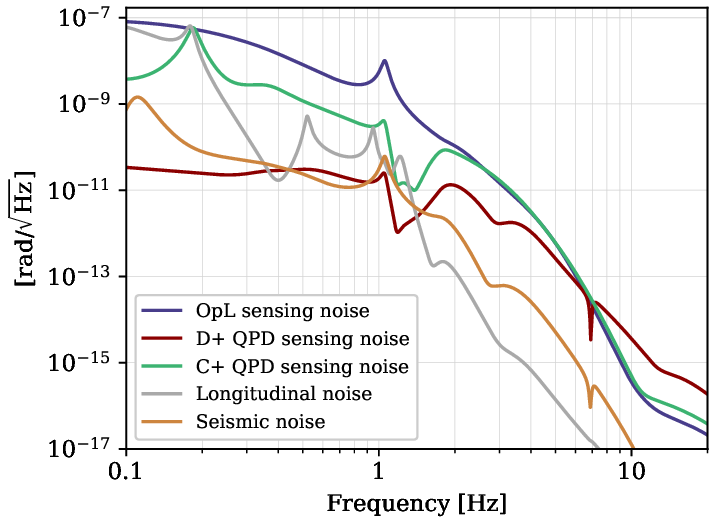}
	\caption{\label{fig:mirror_motion}The figure shows the movement of one of the mirrors, specifically WI, caused by each input noise. Each contribution to mirror motion was calculated using the \finesse~model, with the methodology explained at the beginning of Appendix~\ref{appendix:noise}.}
\end{figure}

The error signals in Fig.~\ref{fig:errors} capture the motion of the DoFs whenever it exceeds the sensor noise. Conversely, when the actual motion of the DoFs is lower than this noise, the error signals consist mainly of sensing noise shaped by the feedback action of the control loops. This noise is then reintroduced into the system, ultimately causing the mirrors to move. Using the model, the actual motion of the mirrors was computed for the given noise sources, and the results are shown in Fig.~\ref{fig:mirror_motion}. From the motion of the mirrors, we calculated the BSM on the TMs using Eq.~\ref{eq:bsm}; the result is shown in Fig.~\ref{fig:bsm}. This is the first time the measured BSM has been accurately reproduced using a numerical model of the optomechanical system.

\section{Noise projections: the methodology}\label{appendix:proj}
\begin{figure}[!t]
	\centering
	\includegraphics[width=\linewidth]{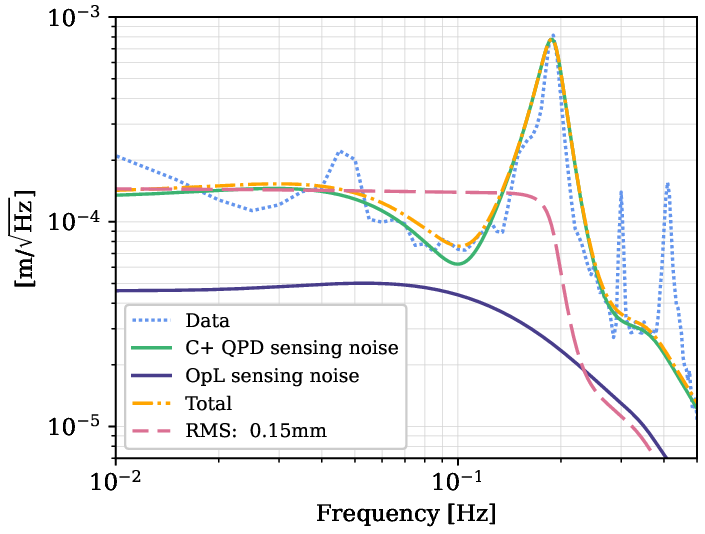}
	\caption{\label{fig:bsm} Noise budget of the beam spot motion. The yellow curve represents the summation of all contributions from the input noise. The dashed red line illustrates the cumulative RMS of the yellow trace, which corresponds to approximately $0.15$~mm. The measured RMS from the experimental data is around $0.14$~mm. Each contribution to the BSM was calculated using the \finesse~model.}
\end{figure}
In Fig.~\ref{fig:block_diagram}, the green branch of the diagram is the chain that carries the motion of the mirrors to the detector output channel. The matrix $d\in\mathbb{R}^{4\times4}$ serves to convert the mirror rotations $\theta_{\text{mir}}$ into the corresponding length changes, resulting in a length vector $z\in\mathbb{R}^{4}$. This matrix $D$ is diagonal and consists of scalar coefficients along the diagonal, as shown below.
\begin{equation}
D = \textnormal{diag}(d_{\textnormal{NI}}  \; d_{\textnormal{NE}}  \; d_{\textnormal{WI}} \;  d_{\textnormal{WE}})
\end{equation}

These coefficients are responsible for translating the rotation of specific optics into length changes. Specifically, when applying the matrix $D$ to the vector $\theta_{\text{mir}}$, the result is the vector $z$, where each element corresponds to a change in the length of the arm cavities due to rotation of each TM:
\begin{equation}
z = [z_{\textnormal{NI}} \; z_{\textnormal{NE}} \; z_{\textnormal{WI}} \;  z_{\textnormal{WE}}]^{T}
\end{equation}
For instance, $z_{\textnormal{NI}}$ represents the modification in the northern cavity's length caused by the rotation of the NI optic. Each element in vector $z$ results from a product of this form:
\begin{equation}
z_{\textnormal{TM}} = d_{\textnormal{TM}}\, \theta_{\textnormal{TM}}
\end{equation}

This relationship resembles the one defined in Eq.~\ref{eq:angle2length0}, where the outcome, in fact, represents a change in length. While $\theta_{\textnormal{TM}}$ encompasses the entire spectrum of NI movement, $d_{\textnormal{TM}}$ has been approximated as a scalar constant, which we call static offset; the same has been done for all other TMs. As explained in~\cite{Barsotti_2010}, while an accurate representation of this process involves the convolution of the BSM spectrum with the angular motion spectrum, the predominance of low-frequency components in the BSM allows us to treat the misalignment as if it were static. The static offset value can be calculated as the RMS of the BSM. The way in which the detector responds to the length variations defined in $z$ is defined by $H \in \mathcal{R}^{4}$, which represents the detector response to a differential change in the lengths of the arm cavities. The dot product between the vectors $H$ and $z$ is the signal produced in the detector output channel $h$, which has been calculated using \finesse.\\

\begin{table}[!b]
    \caption{Arrangement of the static offset on each TM for the projection of each DoF.}
    \vspace{.1cm}
    \label{tab:static_offset}
    \centering
    \small 
    \begin{tabular}{|l|c|c|c|c|}
    \hline
     & \textbf{NI} & \textbf{NE} & \textbf{WI} & \textbf{WE} \\
    \hline
    \textbf{D}$_+$   & $-d_{\textnormal{RMS}}$ & $+d_{\textnormal{RMS}}$  & $-d_{\textnormal{RMS}}$  & $+d_{\textnormal{RMS}}$  \\
    \hline
    \textbf{C}$_+$   & $-d_{\textnormal{RMS}}$  & $+d_{\textnormal{RMS}}$  & $+d_{\textnormal{RMS}}$ & $-d_{\textnormal{RMS}}$  \\
    \hline
    \textbf{N}$_-$  & $+d_{\textnormal{RMS}}$  & $+d_{\textnormal{RMS}}$  & \;\;\;\;\;$0$ & \;\;\;\;\;$0$ \\
    \hline
    \textbf{W}$_-$  & \;\;\;\;\;0 & \;\;\;\;\;0 & $+d_{\textnormal{RMS}}$  & $+d_{\textnormal{RMS}}$\\
    \hline
    \end{tabular}
\end{table}

The amount of angular control noise calculated through the AdV model is shown in Fig.~\ref{fig:projections_tot}. In terms of the methodology used to generate these noise projections, as well as for all the noise projections presented in this work, the process comprises two fundamental steps. First, external disturbances are propagated through the model to calculate the motion of the test masses and subsequently determine the beam spot motion, as shown in Fig.~\ref{fig:bsm}. The RMS of the BSM is then used as the static offset, which in our case has been determined to be $d_{\textnormal{RMS}}=0.15$~mm. Next, this beam offset is applied to all the TMs, and the external disturbances are propagated again through the model to calculate the differential change in the length of the detector's arms. In particular, during this second step, only sensor noise has been propagated because it is the dominant noise in the detection frequency band ($f>10$ Hz), which is now our focus; see Fig.~\ref{fig:errors}. Fig.~\ref{fig:projections_tot} shows the total level of angular noise produced by all the DoFs under control. The contribution of each of the DoF is shown in Fig.~\ref{fig:projections_dof}. Each contribution was obtained by propagating the sensor noise used to control that particular DoF. For example, the C$_+$ curve was obtained by propagating the C$_+$ QPD sensing noise. The last point to consider is how the static offset has been allocated to each TM. The used noise projection methodology takes into account the coherence between the movements of all mirrors. For example, when the C$_+$ QPD sensing noise is propagated, the TMs rotate simultaneously in phase. This results in a noise projection depending on the arrangement of the static offset on each TM. For example, if the beam spot is offset in such a way that the beams in both cavities are shifted, the in-phase motion of the TMs does not change the cavity length and, therefore, does not lead to the generation of angular noise. The static offset has been distributed among the TMs to maximize the coupling of each DoF to the detector output. Each curve in Fig.~\ref{fig:projections_dof} corresponds to a different distribution of the direction of the static offset. For example, in the case of the C$_+$ curve, the matrix $D$ was defined as:
\begin{figure}[!t]
	\centering
	\includegraphics[width=\linewidth]{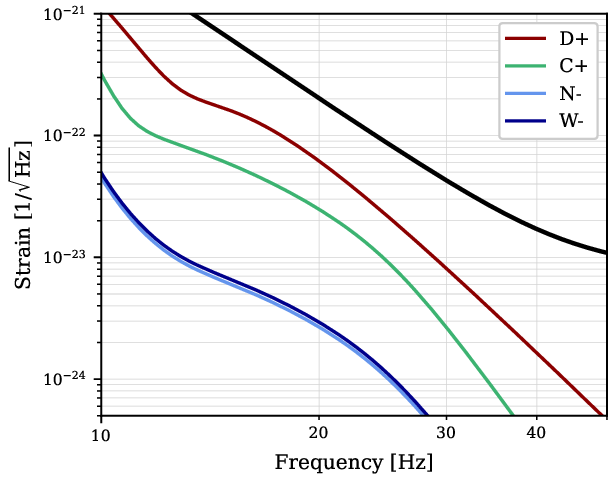}
	\caption{\label{fig:projections_dof}The figure displays the angular control noise generated by each DoF under control. The methodology used to produce this projection is explained in Appendix~\ref{appendix:proj}. The arrangement of the static offsets on each of the TMs is shown in Tab.~\ref{tab:static_offset}. This arrangement ensures that the coupling between the controlled DoF and the detector output is maximized. In black, the target sensitivity for AdV during O3 is shown.}
\end{figure}
\begin{equation*}
D_{\textnormal{C}+} = \textnormal{diag}(-d_{\textnormal{RMS}}  \;\; +d_{\textnormal{RMS}}  \;\; +d_{\textnormal{RMS}} \;\;  -d_{\textnormal{RMS}})
\end{equation*}
More generally, the distribution of the static offset for the projection of each degree of freedom is shown in Tab.~\ref{tab:static_offset}. 

\section{A brief overview of ET mechanics}\label{appendix:ETmech}
In this section, we provide details of the mechanics used for ET. Specifically, we adapted key parameters from the Virgo suspension to ET. All mechanical details considered are shown on Tab.~\ref{tab:et_mech}. Starting with the mass, thickness, and radius of the TM, its moment of inertia was defined using the formula for a solid cylinder:
\begin{equation}
    I_{\text{mir}} = \frac{m\, (3r^2+h^2)}{12}
\end{equation}
The TM is attached to the marionette by four suspension wires. The tension in a single wire can be calculated as:
\begin{equation}
    T_{\text{w}} = \frac{m \,g}{4}    
\end{equation}
The factor of 4 takes into account that the tension is distributed among four wires.
At this point, the stiffness of the final stage of the suspension, defined by the overall stiffness of the four wires, was calculated as follows:
\begin{equation}
    k_{\text{mir}} = \frac{mr^2}{l}
\end{equation}
where $l$ is the length of a wire.\\
As for the marionette, its moment of inertia, we have simply maintained the same marionette-to-mirror proportions as in the Virgo suspension, where the marionette's moment of inertia is approximately four times that of the mirror. Concerning its stiffness, this was manually adjusted to have the resonance frequency of the puppet stage around $20$~mHz, as in the case of Virgo. Regarding the effects of radiation pressure, they are included by default in \finesse~simulations.

\begin{table}[b!]
  \caption{Parameters of the mechanics considered for ET. The parameters for the TM are based on~\cite{Et2020}.}
  \label{tab:et_mech}
  \begin{ruledtabular}
    \begin{tabular}{lc}
      \textbf{Parameter} & \textbf{Value} \\
      \hline
      TM mass $m$ & $211$ kg \\
      TM radius $r$ & $0.225$ m \\
      TM thickness $h$ & $0.57$ m \\
      Wire length $l$  & $0.9$ m \\
      TM stage inertia $I_{\text{mir}}$ & $8.4$ kg·m$^2$ \\
      TM stage stiffness $k_{\text{mir}}$ & $116$ Nm/rad \\
      Marionette stage inertia $I_{\text{mar}}$ & $33.5$ kg·m$^2$ \\
      Marionette stage stiffness $k_{\text{mar}}$ & $0.53$ Nm/rad\\
    \end{tabular}
  \end{ruledtabular}
\end{table}

%
\bibliography{biblio}

\end{document}